\documentclass[prb,twocolumn,showpacs,superscriptaddress]{revtex4-1}

\usepackage{amsmath}
\usepackage{amssymb}

\usepackage{longtable}%
\usepackage{dcolumn}
\usepackage{bm}
\usepackage{color}

\usepackage{graphicx}
\usepackage{dcolumn}
\usepackage{bm}
\usepackage{amsmath}
\usepackage{tabu}
\usepackage{multirow}
\usepackage[latin1]{inputenc}
\usepackage{subfigure}
\usepackage{braket}
\usepackage{soul}
\setstcolor{red}
\usepackage{xcolor}
\usepackage{verbatim}
\usepackage[dvipdfm, pdfstartview=FitH, CJKbookmarks=true, bookmarksnumbered=true,
bookmarksopen=true, colorlinks, pdfborder=001, linkcolor=blue, anchorcolor=blue, citecolor=blue]{hyperref}

\begin{document}
\preprint{Regular article}



\title{Nematic state of the FeSe superconductor}
\author{Sahana R\"o{\ss}ler}
\email{roessler@cpfs.mpg.de}
\affiliation{Max Planck Institute for Chemical Physics of Solids,
N\"othnitzer Stra\ss e 40, 01187 Dresden, Germany}
\author{Mauro Coduri}
\affiliation{University of Pavia, Chemistry Department, via Taramelli 16, 27100, Pavia, Italy}
\author{Alexander A. Tsirlin}
\affiliation{Experimental Physics VI, Center for Electronic Correlations and Magnetism, University of Augsburg, 86159 Augsburg, Germany}
\author{Clemens Ritter}
\affiliation{Institut Laue-Langevin, 71 Avenue des Martyrs, CS20156, 38042 Grenoble C\'edex 9, France}
\author{Gabriel~Cuello}
\affiliation{Institut Laue-Langevin, 71 Avenue des Martyrs, CS20156, 38042 Grenoble C\'edex 9, France}
\author{Cevriye~Koz}
\affiliation{Max Planck Institute for Chemical Physics of Solids,
N\"othnitzer Stra\ss e 40, 01187 Dresden, Germany}
\author{Liudmila Muzica}
\affiliation{Max Planck Institute for Chemical Physics of Solids,
N\"othnitzer Stra\ss e 40, 01187 Dresden, Germany}
\author{Ulrich Schwarz}
\affiliation{Max Planck Institute for Chemical Physics of Solids,
N\"othnitzer Stra\ss e 40, 01187 Dresden, Germany}%
\author{Ulrich K. R\"o{\ss}ler}
\affiliation{IFW Dresden, Helmholtzstra\ss e 20, 01069 Dresden, Germany}
\author{Steffen Wirth}
\affiliation{Max Planck Institute for Chemical Physics of Solids,
N\"othnitzer Stra\ss e 40, 01187 Dresden, Germany}
\author{Marco Scavini}
\email{marco.scavini@unimi.it}
\affiliation{Dipartimento di Chimica, Universita degli Studi di Milano, Via Golgi 19, 20133 Milano, Italy}
%
\date{\today}

\begin{abstract}

We study the crystal structure of the tetragonal iron selenide FeSe and its nematic phase transition to the low-temperature orthorhombic structure using synchrotron x-ray and neutron scattering analyzed in both real and reciprocal space. %
We show that in the local structure the orthorhombic distortion associated with the electronically driven nematic order is more pronounced at short length scales. It also survives up to temperatures above 90~K where reciprocal-space analysis suggests tetragonal symmetry. Additionally, the real-space pair distribution function analysis of the synchrotron x-ray diffraction data reveals a tiny broadening of the peaks corresponding to the nearest Fe--Fe, Fe--Se, and the next-nearest Fe--Se bond distances as well as the tetrahedral torsion angles at a short length scale of 20\,{\AA}. This broadening appears below 20~K and is attributed to a pseudogap. However, we did not observe any further reduction in local symmetry below orthorhombic down to 3\,K. Our results suggest that the superconducting gap anisotropy in FeSe is not associated with any symmetry-lowering short-range structural correlations.

\end{abstract}


\maketitle
\section{Introduction}
The discovery of superconductivity in FeSe \cite{Hsu2008} with a critical temperature $T_c \approx8\,$K prompted intense research in this remarkable family of materials \cite{Boh2018,Kre2020,Shi2020}. 
FeSe, the structurally simplest among the iron-based superconductors, was considered to be an ideal candidate for studying the mechanism of unconventional multiband superconductivity. In spite of the expectations, more than a decade of research proved FeSe as a rather unusual superconductor with several fascinating properties in the normal state, i.e., above $T_c$. 

A structural transition from the parent tetragonal ($P4/nmm$, No. 129) crystal structure to the orthorhombic $Cmma$ (new symbol $Cmme$, No. 67) space group occurring at a temperature $T_s \approx 90$~K is considered to be driven by electronic degrees of freedom, and hence referred to as a nematic transition. However, the energy scale of this nematic order ranges from 10--50\,meV, as determined from angle-resolved photoemission experiments (ARPES), and
is controversially discussed in the literature\cite{Fed2016,Col2018,Yi2019}. At ambient pressure, the nematic transition is not accompanied by a long-range magnetic order \cite{Mc2009a}. Recent reports on pair distribution function (PDF) analysis of scattering data have shown that the short-range orthorhombic distortions are already present at room temperature with a growing domain size upon decreasing temperature \cite{Koch2019,Benj2019,Kon2019,Wang2020}. 

Even though the primary order parameter of the phase transition at 90~K is considered as nematic and of electronic origin, the exact nature of the order parameter remains unknown\cite{Fer2014}. Several possible order parameters have been suggested, which include spin \cite{Wang2016a}, charge \cite{Mass2016}, or orbital \cite{Baek2014} degrees of freedom, antiferroquadrupolar order \cite{Yu2015}, stripe quadrupolar order \cite{Zha2017}, and collective modes such as a Pomeranchuk instability \cite{Mass2016,Kle2018} of the Fermi surface. Indeed, in a quantum material such as FeSe, the order parameters cannot be considered as independent and competing. Instead, the order parameters are composite and intertwined as suggested by recent theoretical studies \cite{Fer2019}.  

Below $T_s$, FeSe displays two more anomalies \cite{Ros2015,Ros2018,Su2016,Ter2016,Imai2009,Gri2018,Kas2014,Ghini2021,Kasa2016} before entering into the superconducting phase at 8~K. The anomaly at $T^* \approx 75$~K was identified in nuclear magnetic resonance (NMR) \cite{Baek2014,Imai2009}, muon spin relaxation ($\mu$SR) \cite{Gri2018}, Kohler's scaling behavior of magnetoresistance, and Hall effect measurements \cite{Ros2015,Ros2018}. Since all these techniques probe the underlying magnetic response of the sample, the temperature $T^*$ is considered as an onset temperature of anisotropic spin-fluctuations. Upon further lowering of temperature to $T^{**}\approx 30-20$~K, an inflection was observed in Hall, Seebeck, and Nernst coefficients \cite{Ros2015,Kas2014,Kasa2016}. The Kohler's scaling was found to be reestablished below $T^{**}$ \cite{Ros2015,Ros2018,Su2016,Ter2016}. 
Although the signature of $T^{**}$ was identified in many different techniques, its interpretation differs in  literature. Several authors propose giant superconducting fluctuations \cite{Kasa2016,Sol2020} or a pseudogap related to preformed Cooper pairs assuming that the system is in the cross-over regime between a weak-coupling Bardeen-Cooper-Schrieffer (BCS) and strong-coupling Bose-Einstein condensate (BEC) limits \cite{Kas2014,Kasa2016}. A breakdown of quantum critical fluctuations owing to a temperature-induced Lifshitz transition was proposed based on the results of $\mu$SR measurements \cite{Gri2018}. A study based on transmission electron microscopy (TEM) suggested 
a possible crystal symmetry lowering below 20~K\cite{Mc2009a}.

  For the mechanism of superconductivity in Fe-based materials, several different pairing scenarios are considered \cite{Kre2020,Maz2008,Kon2010,Islam2021,Fer2022}. Experimental studies based on scanning tunneling microscopy/spectroscopy (STM/S) and ARPES on FeSe have reported highly anisotropic superconducting gaps with either deep gap minima \cite{Lin2016,Spr2017,Kush2018} or existence of nodes \cite{Kas2014,Hashi2018}. Detailed mappings of the Fermi surfaces have found a $\Gamma$--centered hole pocket and an $X$--centered electron pocket with the gap minima along the $k_y$ axis for the $\Gamma$ pocket and along the $k_x$ axis for the $X$ pocket, respectively \cite{Spr2017,Kas2014}. This strong anisotropy in the superconducting gap symmetry has been attributed to orbital selective interactions \cite{Nic2013,Liu2018,Steff2021,Pfau2021,Bart2021}. Moreover, inelastic neutron scattering studies have identified a prominent role of spin-orbit coupling (SOC) in FeSe \cite{Ma2017}, which provides anisotropy to the spin channel. Through the SOC, the spin degrees of freedom couple to the lattice \cite{Lav2002,Ross2019}. 
	
With the motivation to probe the couplings of the electronic degrees of freedom to the crystal structure, we performed synchrotron powder x-ray diffraction and powered neutron scattering experiments in the tetragonal, nematic, and superconducting phases of FeSe. The powder diffraction experiments allow the highest resolution in order to distinguish the possible symmetry lowering below orthorhombic and can be directly converted into pair distribution function as the best tool for probing the local structure.
These experiments enable us to identify if any incipient structural ordering modes are associated with the observed anomalies in the physical properties at different temperatures.     
\begin{figure*}[t]
\centering
 \includegraphics[clip,width=1.95\columnwidth]{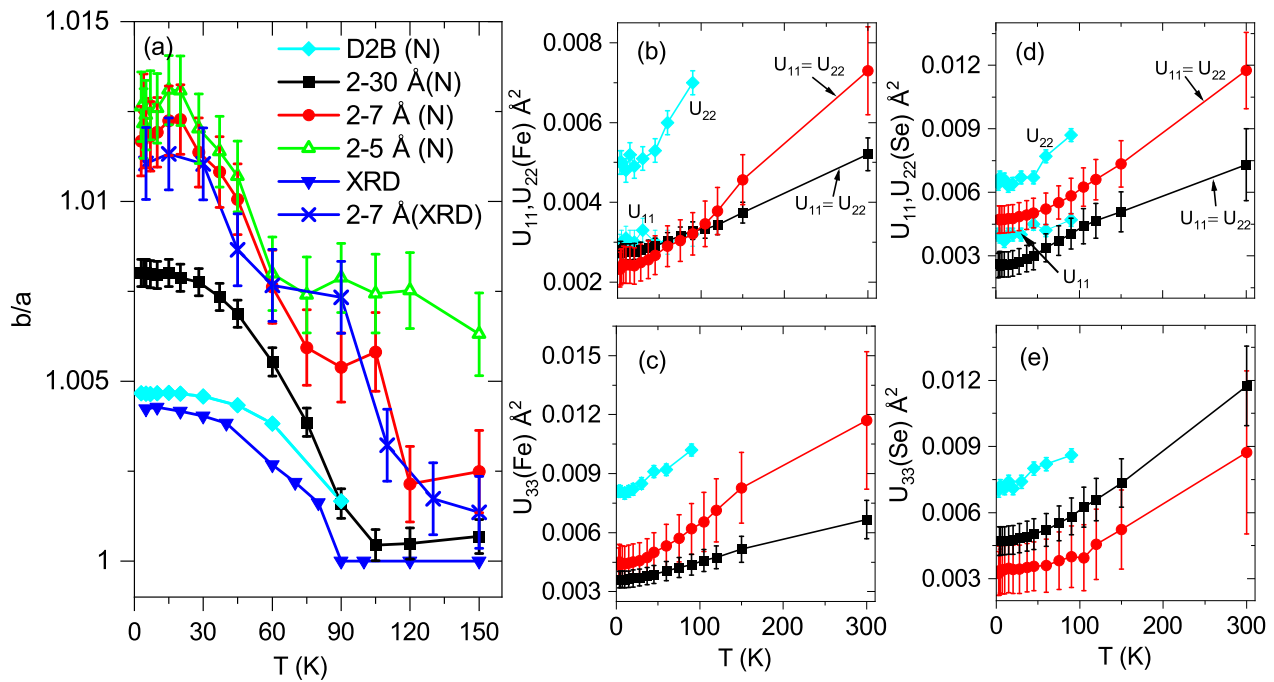}
\caption{A comparison of parameters obtained from the reciprocal-space refinement of neutron (D2B) and x-ray (XRD) diffraction data, as well as real-space analysis of $G(r)$ in different $r$-ranges (using both x-ray and neutron, assigned as XRD and N, respectively in the legend). (a) Ratios of lattice parameter $b$/$a$ and (b)-(e) anisotropic  displacement parameters $U_{11}$, $U_{22}$ and $U_{33}$, obtained from both reciprocal-space refinements of D2B measurements and real-space Rietveld analysis of D4c $G(r)$ functions for different $r$ ranges. In the $G(r)$ refinements, we fixed $U_{11}$\,=\,$U_{22}$ for both Fe and Se. Legend in (a) applies to all panels. The error bars for the D2B and XRD data shown in (a) are smaller than the symbol size.}      
\label{Fig1}
\end{figure*}
%
%
%
%
%
\section{Experimental}
Polycrystalline FeSe samples were synthesized by following the procedure described in Ref.\,[\onlinecite{McQ}].
The samples were characterized using laboratory x-ray diffraction. 

The synchrotron x-ray diffraction was carried out at the beamline ID22 (wavelength $\lambda=0.354144(6)\,\mathrm{\AA}$) of the ESRF (European Synchrotron Radiation Facility, Grenoble) using a special He-flow cryostat or a cold N$_{2}$--gas blower (cryostream) adapted to the diffraction setup. At ID22, 9 scintillation detectors preceded by Si analyzer crystals were used. This allowed an extremely high resolution but also caused long counting times in PDF measurements. The powder sample was loaded into a 0.5 mm glass capillary and closed with a wax to allow a good thermal contact of the sample with an exchange gas. Here we would like to note that the very high photon flux of the ID22 beamline causes local heating of the sample. Hence, the x-ray beam was attenuated by about 70\,\% for our measurements. Further attenuation of the beam did not change the lattice parameters or the temperature of the tetragonal-orthorhombic transition indicating a stable local sample temperature. 

The x-ray diffraction data were collected at temperatures from 5 to 150\,K, and in the $2\theta$ range $ 2^{\circ} \leq 2\theta \leq 90 ^{\circ}$ using the cryostat due to system constraints, and in the range $ 2^{\circ} \leq 2\theta \leq 120^{\circ}$ for measurements using the cryostream. In both cases the scanning rate was $4^{\circ}$/min. For PDF analysis, the data were collected by summing several scans for a total counting time of 385 min per measured temperature to achieve the necessary data quality for analysis. 


The neutron scattering experiments were performed at ILL (Institut Laue Langevin, Grenoble). For these measurements, FeSe powder samples were synthesized in several batches and about 5\,g of impurity free\cite{Koz2014} FeSe powder was selected. The sample was filled in a vanadium container. 

The neutron diffraction experiments were performed in two different instruments. For the reciprocal-space structure analysis (Rietveld refinement) the data were collected on the high resolution powder diffractometer D2B ($\lambda = 1.596\,\mathrm{\AA}$) in the temperature range 3--90\,K in the heating cycle (ILL D2B data Ref.\,[\onlinecite{RefD4data}].) For real-space PDF analysis, the diffraction patterns were collected at the D4c instrument \cite{Fischer2002} ($\lambda = 0.4959\,\mathrm{\AA}$) at temperatures 150 and 7 K upon cooling and in the temperature range 3-150 K plus 298 K, upon heating (see ILL D4c data Ref. [\onlinecite{RefD4data}]).
However, since no hysteresis was found in the structural data, we present here only the data collected in the heating cycle. 

Reciprocal-space refinement of the x-ray and neutron diffraction data was performed using the Rietveld method implemented in GSAS\cite{Gsas} and JANA2006\cite{Jana}. Moderate effects of preferred orientation caused by the large anisotropy of the crystal structure were accounted for by the March-Dollase formalism \cite{Mar1932,Doll1986}. 

The X-ray PDF curves G(r) were computed using the PDFgetX3 program \cite{PDFGetX3}. 
The neutron D4c data reduction was performed using standard instrument-dependent software available at the ILL.  The CORRECT program \cite{Howe1996} was then used to conduct the necessary corrections (see, for example ref.\,[\onlinecite{Fischer2006}]) to the diffraction data for sample and container attenuation, multiple-scattering, and normalization of the diffraction intensity to an absolute scale using a vanadium standard.  The resulting $S(Q)$ data were then Fourier transformed to a PDF using the standard sine integral.

\section{Results}

Here, we would first like to outline a few advantages of using both synchrotron x-ray and neutron scattering experiments for structural analysis. For the reciprocal-space structure analysis, neutron data are often preferred, because the reflection intensity does not go down with $Q=4\pi \mathrm{sin\theta}/\lambda$, so that we can observe a higher number of intense reflections and use them in the structure refinement. Moreover, neutrons offer a broader and more symmetric peak shape, whereas x-rays may show different sorts of anisotropic reflection broadening due to the sample\textsc{\char13}s microstructure. Such anisotropic broadening is undesirable for the structure refinement.  

For the real-space structure analysis, the x-rays and neutrons provide complementary information. With the instruments at the ID22, we reached larger $Q_{max}$ values using both the cryostat ($\approx\,25\,\mathrm{\AA}^{-1}$) and the cryostream ($\approx\,29-30\,\mathrm{\AA}^{-1}$). Additionally, the high $Q$ resolution of ID22 allows to analyze the data up to several nanometers\cite{Dej2018}. In the case of neutrons, the usable neutron wavelengths at the D4c diffractometer are 0.3, 0.5, and 0.7 $\mathrm{\AA}$. However, the neutron flux is too low at $\lambda \approx 0.3\,\mathrm{\AA}$ to get good data for the PDF analysis in a reasonable time. Therefore we used the next shortest available neutron wavelength $\lambda \approx 0.5\,\mathrm{\AA}$, which gives a  $Q_{max} \approx\,23.6\,\mathrm{\AA}^{-1}$. Further, the x-ray measurements are complementary to neutrons in the sense that, for neutrons the weight of $G(r)$ is larger for Fe than Se. For the x-rays, the opposite is true. Thus, by combining the two techniques, we are able to maximize the information about the low temperature structure of FeSe.

Analysis of the interatomic distances made use of the reduced PDF, $G(r)$, which is the sine Fourier transform of the experimental total scattering function, $S(Q)$, defined as \cite{Egami2003,All2012}:

 \begin{eqnarray}
G(r) & = &  4 \pi r \, \nonumber 
[\rho(r)-\rho_0] \\ 
\  & = & \frac{2}{\pi}\,
\int_{Q_{\textrm{min}}}^{Q_{\textrm{max}}} \, Q[S(Q)-1]\sin(Q.r)\;dQ, \nonumber 
\end{eqnarray}
where $\rho(r)$ is the atomic number density function and indicates the probability of finding an atom at a distance $r$ from another atom while $\rho_0$ is the average atom number density. The $G(r)$ function measures deviations from the average atomic density. Provided that the scattering lengths of two atoms involved in a certain $G(r)$ peak take the same sign, a positive peak in the $G(r)$ pattern indicates a range of $r$ values whereby the probability of finding interatomic vectors is greater than that determined by the number density, while the opposite holds for negative $G(r)$ values. 
\begin{figure}
\centering
 \includegraphics[clip,width=0.95\columnwidth]{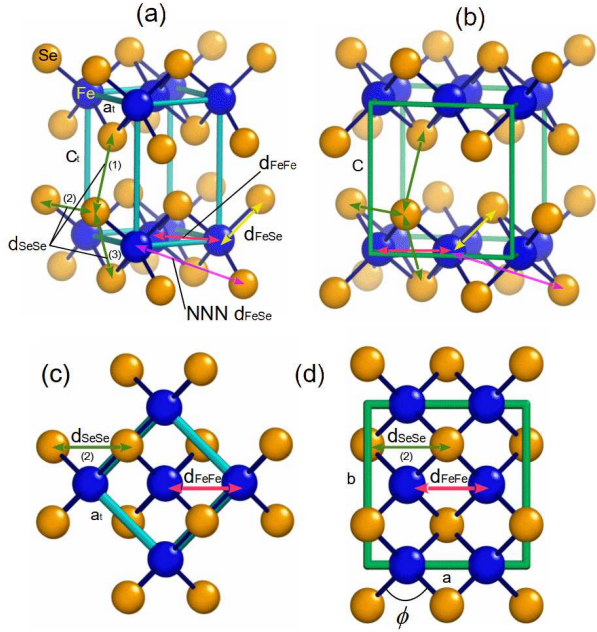}
\caption{Schematic representations of FeSe unit cells in (a) tetragonal and (b) orthorhombic structures. (c) and (d), $ab$ plane of the tetragonal and orthorhombic structures, respectively. The lattice parameters $a$, $b$, $c$, inter-atomic distances, and the torsion angle $\phi$ are marked in the figures. Here, notations $d_\mathrm{FeSe}$ and $d_\mathrm{FeFe}$ represent nearest neighbor, NNN $d_\mathrm{FeSe}$ next-nearest neighbor inter-atomic distances, respectively. To distinguish three different Se--Se distances $d_\mathrm{SeSe}$, numbers (1), (2), and (3) are used, which represent nearest, next nearest, and next-next-nearest neighbor Se--Se, respectively.}      
\label{Fig2}
\end{figure}
\begin{figure}
\centering
 \includegraphics[clip,width=0.95\columnwidth]{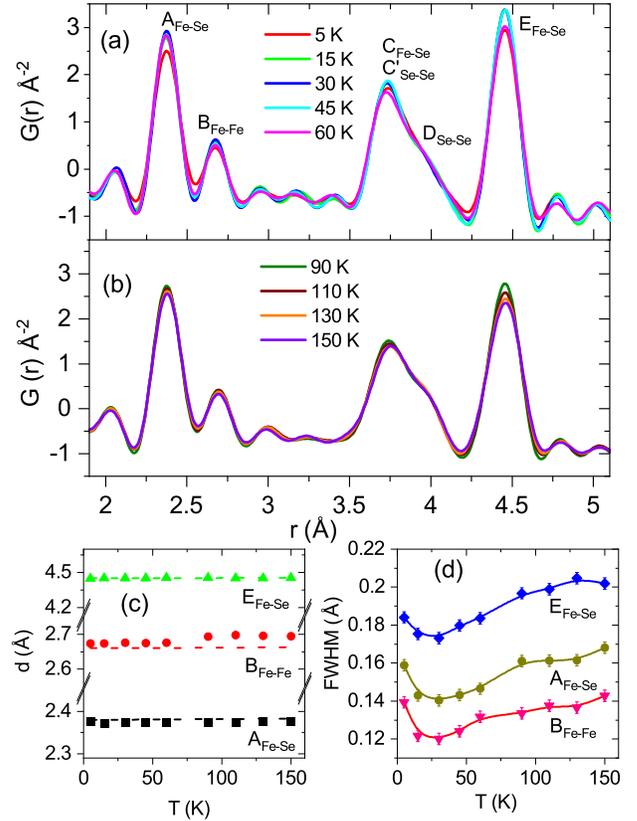}
\caption{(a) and (b) Low $r$ range of $G(r)$ calculated at different temperatures using synchrotron x-ray diffraction. Here, the notations A$_\mathrm{Fe-Se}$, B$_\mathrm{Fe-Fe}$, and C$'_\mathrm{Se-Se}$ represent nearest neighbor, C$_\mathrm{Fe-Se}$ and D$_\mathrm{Se-Se}$ next-nearest neighbor, and E$_\mathrm{Fe-Se}$ next-next-nearest neighbor inter-atomic distances, respectively. (c) The peak positions obtained by the direct analysis of $G(r)$ functions (symbols) and by the reciprocal-space Rietveld analysis (dashed lines). (d) The temperature dependence of full-width half-maxima (FWHM) values obtained from the $G(r)$ peaks displayed in (a) and (b). The error bars in (c) are smaller than the symbol size.}      
\label{Fig2}
\end{figure}

With a focus on investigating local structural distortions associated with the anomalies observed at temperatures $T^*$ and $T^{**}$, we compare the parameters obtained from the analysis of the neutron and x-ray diffraction data for the orthorhombic crystal structure with $Cmma$ symmetry using the so called real-space Rietveld method \cite{PDFGui}. In Fig.\,1(a), the ratio of lattice parameters $b$/$a$ obtained from the reciprocal-space Rietveld analysis (average value of $b$/$a$) and $G(r)$ analysis for different $r$ ranges are plotted. For the x-ray diffraction data, only the average $b$/$a$ ratio and that for the $r$ range 2-7 {\AA} is presented for clarity. Compared to the average values, the $b$/$a$ ratio increases for refinements focusing on the short-range parts of the $G(r)$ functions, which suggests that the orthorhombic strain is locally enhanced. 
In the neutron case, for the 2-7 {\AA} range, the $b$/$a$ ratio displays a plateau in the temperature range 75-105\,K with the extent of this plateau reaching up to 150\,K when the $r$ interval is the shortest. A similar tendency is also seen in the XRD data for the 2-7 {\AA} range, but in this case there are insufficient data points to determine the temperature range of the plateau. Note that 75\,K corresponds to the temperature $T^*$, which is considered as the onset temperature of spin fluctuations \cite{Ros2015,Ros2018,Gri2018}.  

We should also note that in the 2-5 {\AA} range, the $G(r)$ samples mainly distances between atoms in the same FeSe plane while for larger $r$ ranges inter-plane contacts have increasing weight in the $G(r)$. This result suggests a larger crystallographic coherence of the orthorhombic distortion in the $ab$ plane compared to the direction along the $c$ axis. Consistently, the residuals are much smaller in the 2-5 {\AA} fit than in the 2-7 {\AA} range (e.g. at 90 K, $R_{w}$ is 0.029 and 0.059, respectively, for the neutron case).
For the $r$ ranges 2-7\,{\AA} and  2-5\,{\AA} a kink can be seen around 20\,K, which coincides with the electronic anomaly observed at $T^{**}$. Figs.\,1 (b)-(e) display the anisotropic displacement parameters U(Fe) and U(Se) refined in different $r$ ranges. To avoid correlations among the parameters, we assumed $U_{11}=U_{22}$ for both Fe and Se in the $G(r)$ refinements. The anisotropic displacement parameters from the reciprocal space refinement are systematically higher than those obtained from the refinement of the real-space. This difference is well seen in both $U_{11}$ and $U_{33}$ and suggests that displacements of atoms with respect to the average structure occur along and perpendicular to the FeSe planes.

\begin{figure}
\centering
 \includegraphics[clip,width=0.95\columnwidth]{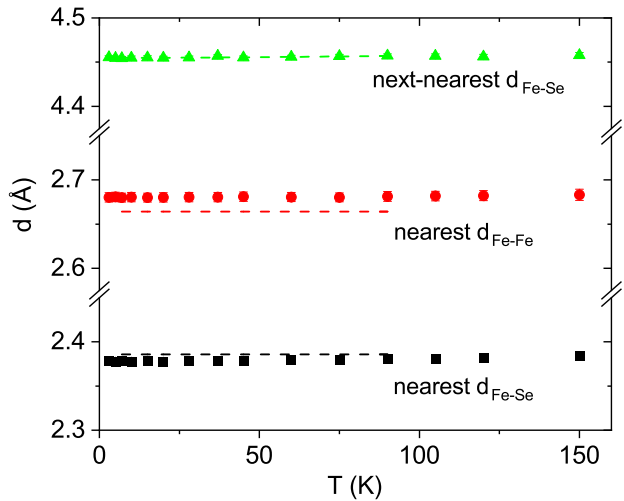}
\caption{Neutron diffraction $G(r)$ peak positions (symbols) in the temperature range 3-75\,K compared with the corresponding distances obtained from the reciprocal-space Rietveld refinement of the D2B data. The error bars are smaller than the symbol size.}      
\label{Fig2}
\end{figure}

\begin{figure*}
\centering
 \includegraphics[clip,width=1.95\columnwidth]{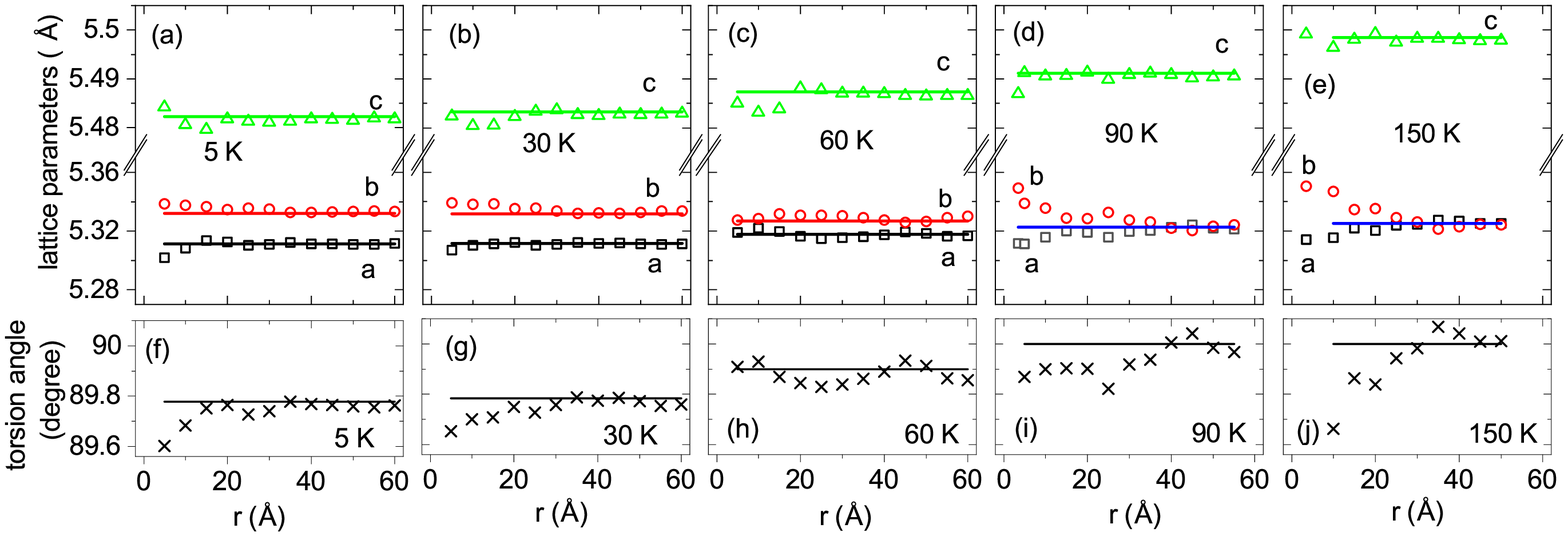}
\caption{(a)-(e) Lattice parameters at different temperatures obtained from $G(r)$ as a function of $r$, using 10 {\AA} wide $r$ intervals. The $r$ values are the centroids of the range. Straight lines are the lattice parameters obtained from the reciprocal-space Rietveld analysis. Blue lines are used in panels (d) and (e) because $a\,=\,b$ in the tetragonal phase. (f)-(j) Torsion angle of the FeSe$_{4}$ tetrahedra as a function of $r$ at different temperatures. The values obtained from the PDF (crosses) replotted as a function of the $r$ range considered and compared to the Rietveld results (black lines).}      
\end{figure*}

In the following, we describe the results of the real-space structural analysis of synchrotron x-ray diffraction data in the temperature range 150--5\,K. To visualize the local distortions described below, in Figs.\,2(a) and (b), we display schematic diagrams of FeSe unit cells in the tetragonal and orthorhombic structures, while Figs.\,2(c) and (d) show the corresponding structures in the $ab$ plane. The figures depict the lattice parameters $a$, $b$, $c$, the inter-atomic distances, and the torsion angle $\phi$ of the FeSe$_4$ tetrahedra. 
In Figs.\,3(a) and (b), the short-range part of the $G(r)$ curves at different temperatures are presented. Due to the wavelength used and the geometrical constraints of the cryostat ($2\theta_{max}=90^{\circ}$) only the data up to $Q_{max}=25\,\mathrm{\AA}$  have been used. The peaks below $r\,<\,3\,\mathrm{\AA}$ and the one around 4.5\,{\AA} have straightforward attributions to single Fe--Se and Fe--Fe inter-atomic distances (see labels in the figure). Conversely, the peaks in the 3.3--4.0\,{\AA} range contain three different Se--Se contributions plus one from Fe--Fe. In all cases, peak broadening is apparent upon lowering temperature. At 5\,K, the peaks A$_\mathrm{Fe-Se}$, B$_\mathrm{Fe-Fe}$ and E$_\mathrm{Fe-Se}$ are broader than at 15 and 30\,K, which is contrary to the expected structural effect when the $G(r)$ peak width arises only from the thermal atomic vibrations. The C$_\mathrm{Fe-Se}$, C$'_\mathrm{Se-Se}$, and D$_\mathrm{Se-Se}$ peaks display some non-monotonic behavior, too. However, due to the superposition of different contributions, it is more difficult to attribute this finding to certain single inter-atomic distances. 

The $G(r)$ peaks shown in Fig.\,3 (a) and (b) have been first analyzed using the so called direct analysis. Each inter-atomic distance has been fitted using a Gaussian function allowing the area, position, and full width at half maximum (FWHM) parameters to vary. The results are summarized in Figs. \,3(c) and (d). In Fig.\,3(c), the peak positions are reported, while in Fig.\,3(d), the relative FWHM values are shown. 

All the FWHM values increase upon cooling only for $T\leq\,15$\,K. However, the change in FWHM $\Delta=$[FWHM(5\,K)- FWHM(15\,K)] are less than 0.02 {\AA}. 
The nearest neighbor Fe-Fe distance (B$_\mathrm{Fe-Fe}$) displays a comparatively larger value of $\Delta\approx 0.018$\,{\AA} than the nearest neighbor Fe-Se (A$_\mathrm{Fe-Se}$, $\Delta\,\approx 0.016$\,{\AA}) and next-next-nearest neighbor Fe-Se (E$_\mathrm{Fe-Se}$, $\Delta\,\approx 0.007$\,{\AA}). Together with the atomic distances obtained from the analysis of $G(r)$ (local structure), the corresponding distances acquired from the reciprocal-space Rietveld analysis (average structure) are shown in Fig.\,3(c) for comparison. For all the distances considered, it can be noted:
(i)	The shortest Fe--Se distance in the local structure is slightly shorter than the corresponding distance in the average structure.
(ii) The shorter Fe--Fe distance in the local structure is always larger than that in the average structure. 
(iii) The Fe--Se distance of 4.4\,{\AA} is the same in both local and average structures.
This trend is confirmed also by the real-space PDF and reciprocal-space Rietveld analysis of the neutron diffraction data, see Fig.\,4.

The very high angular resolution of the ID22 data allows calculating reliable $G(r)$ functions up to tens of angstr\"om. Thus, we analyzed the experimental $G(r)$ functions in the temperature range 5-150\,K assuming the $Cmma$ lattice structure at different $r$ intervals (box car refinements).
Box car refinements were carried out on $r$--intervals 10\,{\AA} wide.
%
%
%
%
At all measured temperatures, no marked deviations from the $Cmma$ structural model are observed. 
An interesting result of the box car refinements is the trend of the lattice parameters, as a function of $r$, as reported in Fig.\,5 (a)-(e), marked by symbols. In the same figure, reciprocal-space Rietveld results are also given for comparison (dashed lines). The orthorhombic strain $\propto\mid{a-b}\mid$ seems to be larger than the average strain at low $r$ values and it reaches the latter value within 20-30\,{\AA}. We would like to note that, for $T\,>\,90$\,K, i.e., above $T_s$, large in-plane orthorhombic strain coexists with weak inter-plane one. For this reason, in the refinements at the lowest $r$ values and $T\,\geq\,90$\,K shown in Fig.\,5, two not matching orders coexist, which leads to less stable refinements.
%
%

The orthorhombic strain is related to the torsion angle of the FeSe$_{4}$ tetrahedra, $\phi =2\mathrm{tan}^{-1}(a/b)$, as defined in Ref.\,[\onlinecite{Mc2009a}]. It is a coherent twisting of the upper and lower Se pairs that make up the Fe-Se tetrahedron; the smaller value of the angle $\phi$ corresponds to the larger orthorhombic strain.
In Fig.\,5(f)-(j), the torsion angles $\phi$ are plotted as a function of $r$. For almost all temperatures up to 150 K, at low $r$, the $\phi$ values are smaller than those in the average structure shown by the straight lines. This confirms our findings from the real-space analysis of the neutron data. However, within a few nm, the $\phi$ values merge into the average angle at all temperatures, and above 90\,K the average structure appears to be tetragonal.


 Remarkably, although the nematic phase transition occurs in FeSe at $T_s\,\approx\,$90\,K, at which the average crystal structure changes from tetragonal to orthorhombic phase, recent PDF studies of FeSe\cite{Koch2019,Benj2019} have shown that the short-range orthorhombic distortion exists already at 300\,K and extends over a length scale of 10-30\,{\AA}. We considered both $Cmma$ and $P4/nmm$ models for the full temperature range studied here. For the neutron diffraction data taken at 298\,K, the $Cmma$ model provided lower residual values than the $P4/nmm$ for $r$ values as small as 5\,{\AA}.
Our results are thus in full agreement with the earlier PDF work in Ref.\,[\onlinecite{Koch2019}]. 

Finally, with the intention to investigate whether there is any deviation from $Cmma$ symmetry at low temperatures, we considered additional models for the PDF analysis. Based on their electron diffraction studies on Fe$_{1.01}$Se McQueen $et\,al.$,\cite{Mc2009a} proposed that the crystal structure at 11\,K is likely of lower symmetry than that of the $Cmma$ space group. They put forward two distortion modes of the in-plane Fe lattice. Either there are alternating displacements of Fe in direction of the short $a$-axis, which may be driven by metal-metal dimer formation, or there are displacements of Fe nearest neighbour-pairs along the long $b$-axis direction. In the latter case, the Fe move alternatingly perpendicular to the distance vector of the pairs that is oriented along the short $a$-direction. This mode avoids dimer formation, but rotates the pairs while keeping their distance.
For both modes, the distortions preserve the C-centering, but remove the glide plane. Combinations or partial (staggered) activation of these two distortion modes would display dimerized and rotated Fe-pairs. Such more complex lattice distortions simultaneously break both the glide mirror and the C-centering \cite{Mc2009a}.
These distortions are subtle and were not observed in the synchrotron x-ray diffraction reported in Ref.\,[\onlinecite{Mc2009a}]. Our PDF analysis shows that these in-plane distortions vanish in the $r$ range of 2--7\,{\AA} and both the models do not provide any improvement in the quality of the fit. We also verified the effect on the refinement of some other possible distortions such as three different types of buckling in which, (i) the Fe $z$ coordinate is shifted up and down forming a checkerboard, (ii) the Fe $z$ coordinate is shifted up and down forming stripes running along $y$, and (iii) the Fe $z$ coordinate is shifted up and down forming stripes running along $x$. Also, a structural model with a $1\times1\times2$ supercell formation was tried. However, the quality of the corresponding fits did not support these scenarios.\\

For the purpose of correlating the electronic properties of FeSe with the structural details determined in this study, we plot in Fig.\,6 the resistivity $\rho(T)$ and the Hall coefficient $R_H(T)$, taken from the Refs.\,[\onlinecite{Lin2016}] and [\onlinecite{Ros2015}]. As summarized in the figure, above the structural transition temperature $T_s$, the average structure is tetragonal, but local incoherent orthorhombic distortions were observed. However, below $T_s$, although electronic anomalies are visible in Fig.\,6 at $T^{*}$ and $T^{**}$, reduction in symmetry below $Cmma$ was not found even at shorter length-scales. Nonetheless, on a local scale, some strain in the lattice was observed as small anomalies in the temperature dependence of the $b/a$ ratio and inter-atomic distances. The average structure remains orthorhombic down to the lowest measured temperature of 3\,K.\\   

\section{Discussion}   

Our structural study involving both reciprocal-space and real-space analysis of the scattering data indicate that the temperature evolution of the static FeSe structure from room temperature through the nematic phase transition at $T_s$ and through further anomalies at lower temperatures is smooth.
Within the experimental resolution, the structural phase transition 
from the tetragonal to the orthorhombic symmetry is 
the only transformation in the system. It leads to a macroscopic
symmetry breaking well above the transition to the superconducting state.
This transition has peculiar features, as seen from the fact that orthorhombic distortions are present in the form of nanodomains already above $T_s$.
Further, as discerned by the $b/a$ ratio in Fig.\,1, on a short length scale (especially at $r=2-5$\,\r A that mainly samples the intraplane interatomic distances), the orthorhombic deviation from tetragonal symmetry is always much stronger than the average value. This behavior suggests that the orthorhombic ordering is highly anisotropic, having a larger coherence length within each FeSe layer than along the $c$ direction.
\begin{figure}[t]
\centering
 \includegraphics[clip,width=0.95\columnwidth]{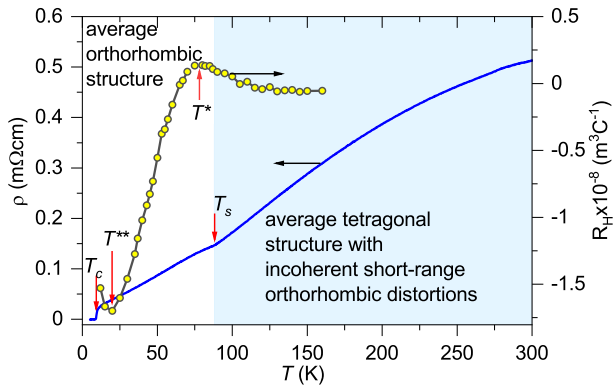}
\caption{Resistivity $\rho(T)$ and the Hall coefficient $R_{H}(T)$ of a FeSe single crystal replotted from the Refs. [\onlinecite{Lin2016}] and [\onlinecite{Ros2015}], respectively. The temperatures $T_s$, $T^{*}$, $T^{**}$, and $T_c$ represent the temperatures of the structural phase transition, onset of spin-fluctuations, onset of the pseudogap, and the superconducting transition temperature, respectively. In our studies, we observe a plateau in the $b/a$ ratio at $T^{*}$, and some broadening in the peaks corresponding to  
Fe--Fe, Fe--Se, and the next-nearest Fe--Se bond distances at $T^{**}$, but the average structure remains orthorhombic below $T_s$.}      
\end{figure}
Above $T_s$,
the orthorhombic domains show coherence lengths of a few nm within the $ab$-planes, but almost no coherence along the $c$-axis. From the perspective of chemical bonds and the electronic structure of FeSe, such a behavior is plausible, as the microscopic origin for the nematic ordering likely pertains to electronic correlations within individual FeSe-layers. A weak coherence along the $c$-direction is also plausible, as the cohesion of the FeSe-layers has a different character with prominent contributions of the van-der-Waals bonding \cite{Ricci2013,Loch2021}. Therefore, the ordering into a three-dimensional long-range order may evolve first through an electronic instability within the FeSe-planes that is already present at high temperatures\cite{Sko2018}. The concomitant square-to-rectangle strains in the $ab$-planes of the tetragonal lattice cannot grow to macroscopic length due to the lack of coherence between the FeSe-planes. Such a mixture of local rectangular lattice patches in the absence of coherence in $c$-direction will appear tetragonal in the average crystal structure. The transition to the long-range ordering at 90\,K then requires an anisotropic divergent growth of domains in the $ab$ plane and possibly a different growth rate of coherence with decreasing temperature in the $c$-direction, as indicated in a recent NMR experiment\cite{Wei2021}. 

Our analysis of the difference between the local orthorhombic distortion measured in real space and the average lattice parameters $a$ and $b$ obtained in the reciprocal space support such a picture of an anisotropic transformation into nematic long-range order. However, there is an additional feature. The data in Fig.\,5 display a tendency to local orthorhombic distortion $b/a$ larger than the average on short lengths below 5\,{\AA} even below  $T_s$. This may indicate that the microscopic nematic instability is frustrated by some additional mechanism and that the nematic state retains a partially incomplete or incoherent character at low temperatures. Further, as there are no important modifications of the lattice structure discernible at and below the transition to the superconducting state at 8\,K, effects of superconducting ordering on the lattice or modifications of the nematic state are not observed at least down to 3\,K.\\

%

\section{Conclusions}
We presented a comprehensive structural analysis of FeSe in the temperature range of 298--3\,K using synchrotron x-ray and neutron diffraction measurements.
Although the global reduction in symmetry from $P4/nmm$ to $Cmma$ was observed only below 90\,K, short-range orthorhombic correlations are found already
at room temperature, which is consistent with previously published results \cite{Koch2019,Benj2019,Kon2019,Wang2020}. No changes in the short-range correlations
were observed when passing through the nematic phase transition from high temperature. A plateau in the $b/a$ ratio was observed at short-length scales around $T^{*}$. 
At the onset temperature of a putative pseudogap $T^{**}$, some broadening in the peaks corresponding to  
Fe--Fe, Fe--Se, and the next-nearest Fe--Se bond distances, and tetrahedral torsion angles were found in the PDF analysis of x-ray diffraction. Our studies show that the local symmetry remains $Cmma$ also in the superconducting state down to at least 3\,K. The local orthorhombic symmetry spanning three orders of magnitude in temperature is yet another trait of FeSe.

\begin{acknowledgments}

We thank ESRF and ILL for granting beam time. We are grateful to Catherine Dejoie and Andy Fitch for providing excellent technical support during our measurements at ID22, ESRF.  
We thank U. Nitzsche for technical assistance.

\end{acknowledgments}

%
%
%
 


\end{document}